\documentclass[prb,aps,floatfix,twocolumn]{revtex4}
\usepackage{epsfig}
\usepackage{amsmath}
\usepackage{amsfonts}
\usepackage{amssymb}
\usepackage{graphicx}
\begin{document}
\title{CeMnNi$_{4}$: an impostor half-metal}
\author{I.I. Mazin }
\affiliation{Center for Computational Materials Science, Naval Research Laboratory,
Washington, D.C. 20375}

\begin{abstract}
Recent experiments show CeMnNi$_{4}$ to have a nearly integer magnetic
moment and a relatively large transport spin polarization, as probed by Andreev
reflection, suggesting that the material is a half metal or close to it.
However, the calculations reported here show that it is not a half metal at all,
but rather a semimetal of an unusual nature. Phonon properties should also be
quite unusual, with rattling low-frequency Mn modes. Nontrivial transport
properties, including a large thermolectric figure of merit, $ZT,$ are
predicted in the ferromagnetic state of the well ordered stoichiometric
CeMnNi$_{4}$

\end{abstract}

\maketitle
%\begin{multicols}{2}

Recently, Singh \ $et\ al$\cite{1} have measured the   magnetic and
transport properties of a novel ferromagnetic material, CeMnNi$_{4}.$ The most
striking observations are that the measured magnetic moment is 4.94 $\mu_{B}%
/$formula, remarkably close to an integer magnetization of 5 $\mu_{B},$ and
at the same time Andreev reflection is suppressed in a way typical of highly
polarized ferromagnets. The degree of spin polarization, deduced in the
standard manner, was up to 65\%, a relatively large number. These observations
together suggest that CeMnNi$_{4}$ might be a half-metal. On the other hand,
another, less obvious, observation cast doubt on such a simple interpretation:
the resistivity as measured in Ref. \onlinecite{1} rapidly grows from zero
temperature to $T_{C}=148$ K,  at a rate up to 2 $\mu\Omega\cdot$cm/K, characteristic
of bad metals, with a very large residual resistivity of 0.24 m$\Omega\cdot
$cm. At the same time, above $T_{C}$ the temperature coefficient of the
resistivity drops practically discontinuously to a value smaller than
0.06 $\mu\Omega\cdot$cm/K, a 1.5 order of magnitude change! Indeed, such
large changes in the temperature coefficient resistivity at $T_C$ have been
previosly encountered only near a metal-insulator transition ($cf.$
collosal magnetoresistance, CMR). Some half metals may exhibit
large changes of the resistivity slope near $T_C$ without 
a metal-insulator transition, but the change is in the opposite direction \cite{CoS2}. 

Band structure calculations for this material can be expected to shed some
light on the puzzling features described above. They do indeed, and in a rather
unexpected way. In this paper I report such calculations and discuss their ramifications.

\begin{figure}[tbp]
\centerline{\epsfig{file=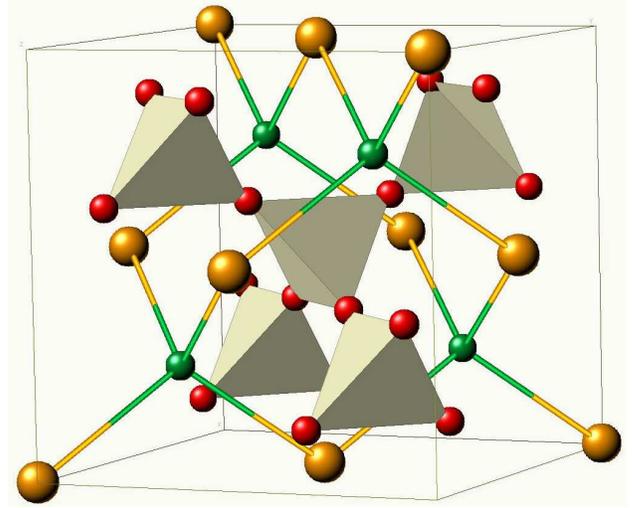,width=0.95\linewidth,angle=0,clip=}}
\caption{Crystal structure of CeMnNi$_{4}$. Large brown spheres denote
the Ce atoms, the small green ones Mn, and the tetrahedra are formed by 
the Ni atoms, denoted by the small red spheres.
(color online) }
\label{struct}
\end{figure}

CeMnNi$_{4}$ crystallizes in the $F\overline{4}3m$ group (\#216). Its
structure can be derived from the Heusler structure $ABCD,$ where Ce and Mn
occupy $A$ and $B$ positions, and Ni sits between $C$ and $D$ (plus three
symmetry equivalent positions), Fig.\ref{struct}. As one can see, Ni forms
corner-sharing tetrahedra, similar to the 
spinel structure. The structure has one
free parameter, the Ni position. If this position is exactly equal to
(5/8, 5/8, 5/8) the lengths of the Ni-Ce and Ni-Mn bonds are exactly the same. As
we will see, the optimized structure is very close to this, despite the fact
that Ce has about 30\% larger atomic radius than Mn. This is yet another hard to
understand property of this compound. I have performed full-potential LAPW
calculations, using the WIEN package\cite{WIEN} and Perdew-Burke-
Ernzerhof\cite{PBE}
gradient-corrected exchange-correlation potential. Muffin-tin radii
of 2.5 $a_{B}$ for Ce and Mn and 2.23 $a_{B}$ for Ni were used, the basis set included
planewaves up to $RK_{\max}=7$ with APW local orbitals, and integration in
 $k-$space was performed using the tetrahedron method with 286 inequivalent
points (21x21x21 mesh)\cite{note}.

\begin{figure}[tbp]
\centerline{\epsfig{file=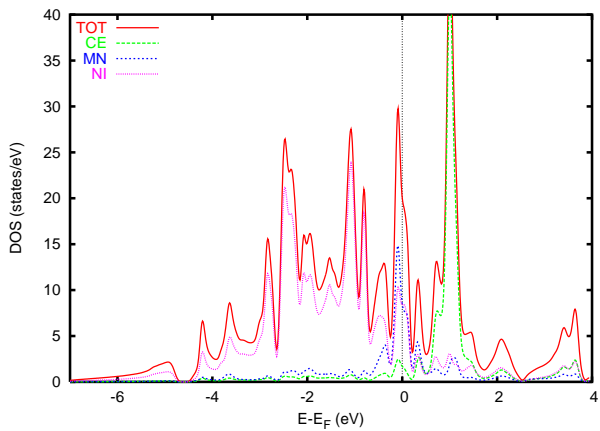,width=0.95\linewidth,angle=0,clip=}}
\caption{Density of states of nonmagnetic CeMnNi$_{4}$.
(color online) }
\label{NONMAG}
\end{figure}

\begin{figure}[tbp]
\centerline{\epsfig{file=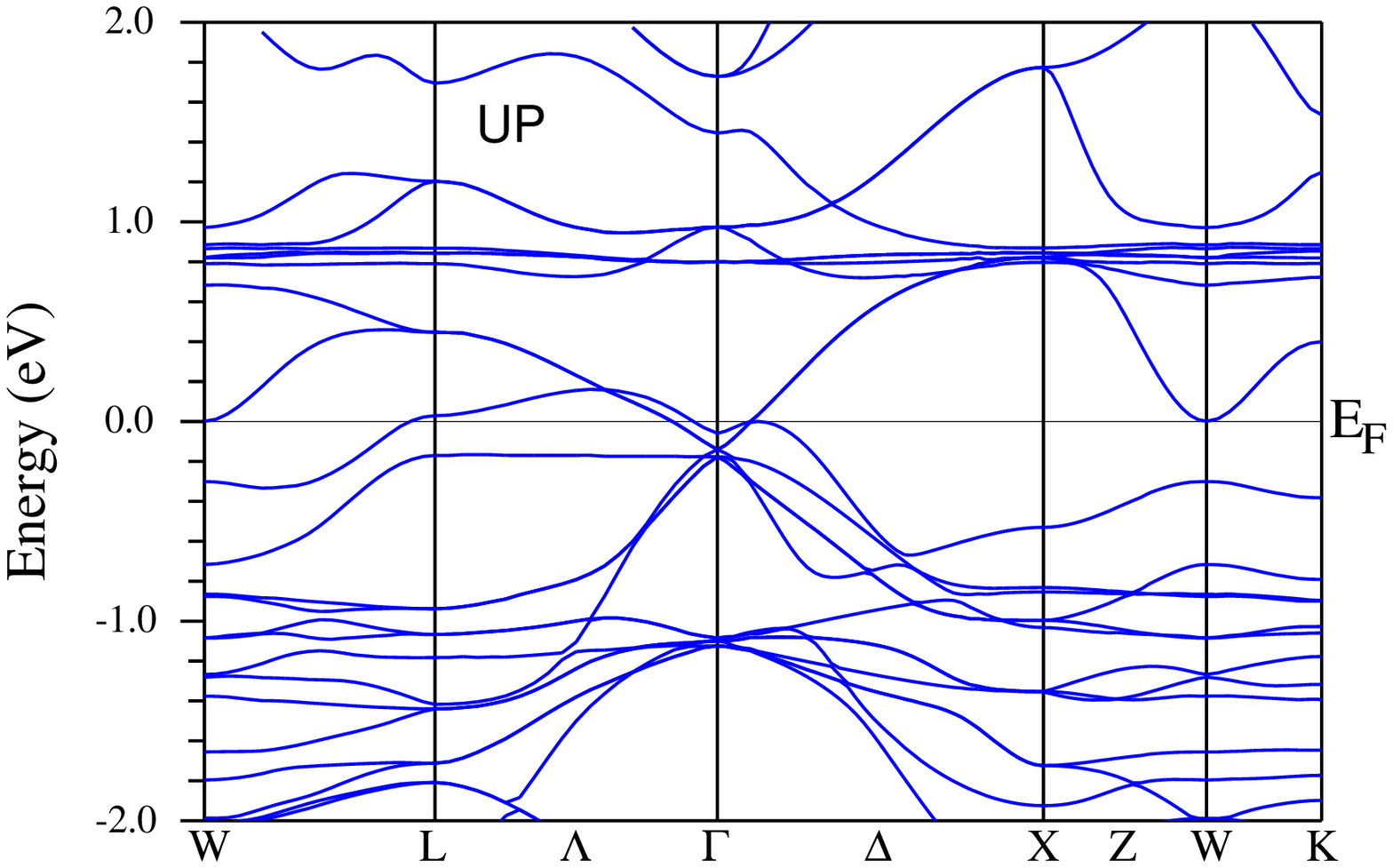,width=0.95\linewidth,angle=0,clip=}}
\centerline{\epsfig{file=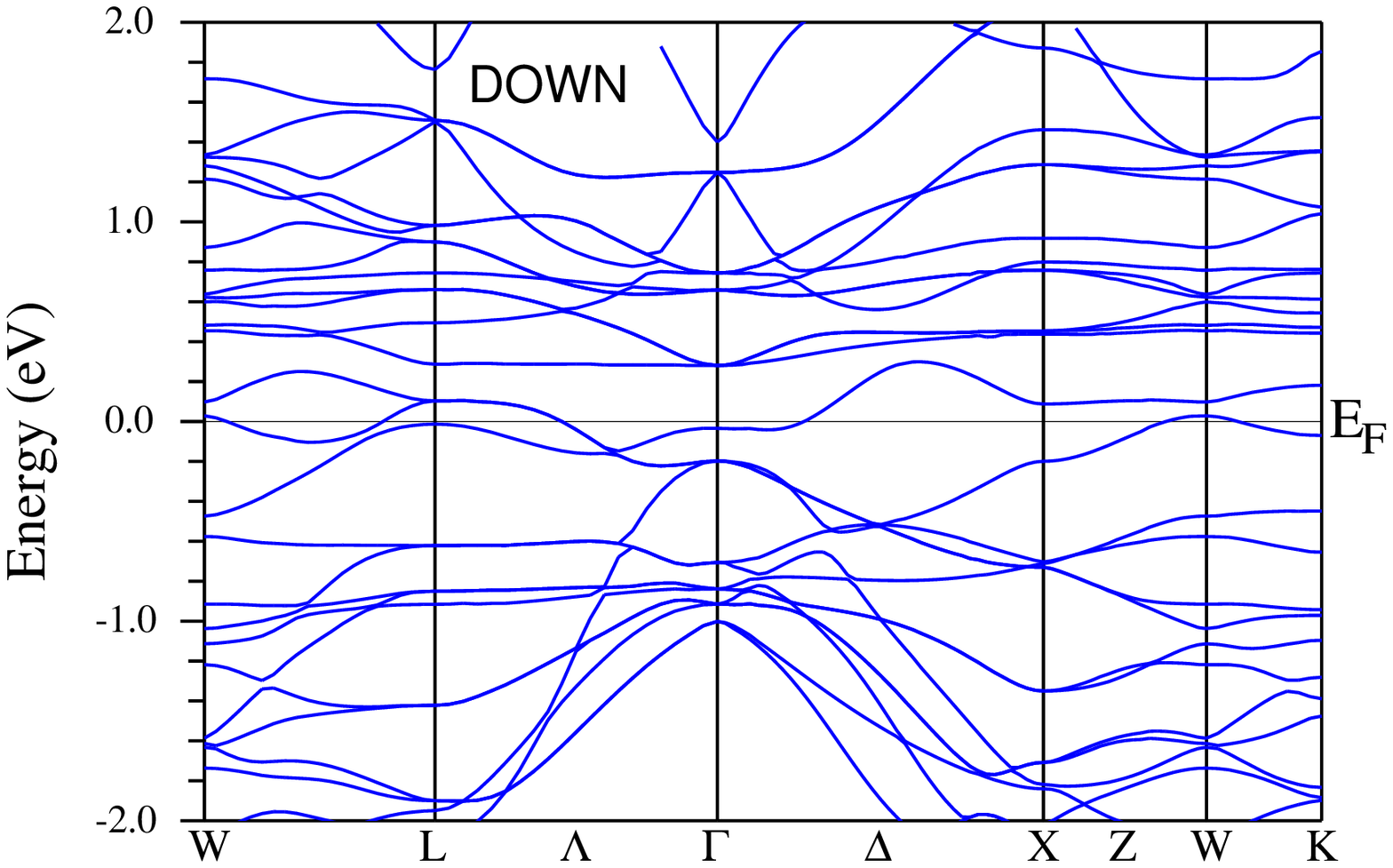,width=0.95\linewidth,angle=0,clip=}}
\caption{Band structuure of the ferromagnetic CeMnNi$_{4}$ in the optimized structure.
Top panel: spin up. Bottom panel: spin down.
 }
\label{OPTbs}
\end{figure}

\begin{figure}[tbp]
\centerline{\epsfig{file=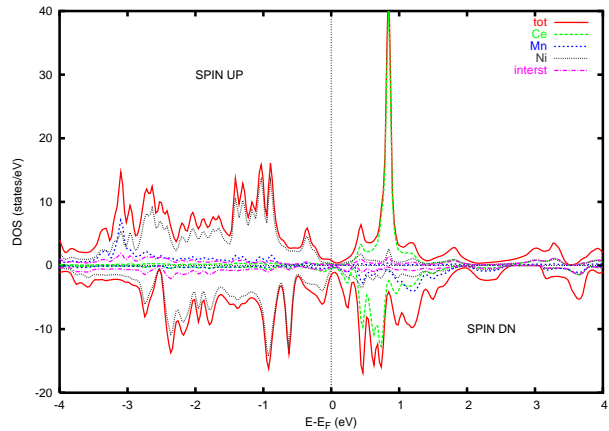,height=0.95\linewidth,angle=-90,clip=}}
\caption{Density of states of the ferromagnetic CeMnNi$_{4}$ in the optimized structure.
 (color online)}
\label{OPTdos}
\end{figure}

The nonmagnetic density of states (DOS) of CeMnNi$_{4}$ is shown in
Fig.\ref{NONMAG}. One can clearly see that Ce f bands are about 1 eV above the
Fermi level, indicating their delocalized character with no need of applying
Hubbard-type correction (\textit{e.g., }within LDA+U). It is further seen that
Mn forms a relatively narrow band (0.25-0.30 eV), while the Ni bands are at
least 4 eV wide (I will explain the origin of the Mn band narrowing later).
Moreover, the
 Mn bands are pinned to the Fermi level, and are largely responsible
for the very high DOS at the Fermi level (10 states/eV.spin.formula, or 2
states/eV.spin per 3d metal ion). Recalling that 3d transition metals have
Stoner factors of the order of 1 eV, it is obvious that even after diluting
with the less magnetic Ce the material should be very strongly magnetic. I thus
proceed with magnetic calculations and find the band structure shown in Fig.
\ref{OPTbs} and \ref{OPTdos}. First, the ferromagnetic structure
is found to be
stabilized by a huge energy gain of 1.87 eV per formula. Second, the total
calculated magnetization is 4.92 $\mu_{B}$/formula, in nearly perfect
agreement with the experiment, and indeed very close to an integer value. The
moment
is distributed like this: Mn carries approximately 4 $\mu_{B}$, four Ni
together about 1.2 $\mu_{B}$ and Ce is polarized antiferromagnetically with a
moment of 0.2 $\mu_{B}.$ Clearly the magnetic engine in this compound is Mn,
whose $d-$states are fully split by about 3 (!) eV\cite{noteU}.
 Ce plays the role of a
cation in this compound, donating its one $f$-electron to Mn. This can be
verified by taking the charges inside each MT sphere and distributing the
interstitial charge proportionally to the MT sphere volumes, which yields
$Q_{Ce}\approx1.2e,$ $Q_{Mn}\approx-0.6e,$ $Q_{Ni}\approx-0.15e.$ As a result,
Mn has 6 $d$-electrons, and full exchange splitting on Mn site results in 5
spin-up and one spin-down electron. Ce $f$ (and $d)$-states are above the
Fermi level, so they hybridize more with the higher-lying 3$d$ metal
spin-down states (mostly Ni) than with the spin-up states, and the former
acquire more of Ce character. This explains the antiferromagnetic polarization
on Ce.

At this point it is worth mentioning that all calculations described above and
below were performed in the structure obtained after optimizing the positions
of Ni by minimizing the total energy in the ferromagnetic state. It appears
that the optimal position of Ni in lattice coordinates is
(0.624, 0.624, 0.624), and symmetry equivalent positions. This is spectacularly
close to the \textquotedblleft ideal\textquotedblright\ position of
(5/8, 5/8, 5/8). Moreover, the corresponding $A_{1g}$ phonon of Ni does not
appear to be particularly soft - the calculated frequency is about 165
cm$^{-1},$ a very regular number for an intermetallic compound with 3d metals.
If one substitutes Mn by Ce, the resulting structure, provided that Ni
occupies the ideal position above, is the well known Laves phase. In fact,
such a phase (CeNi$_{2}$) does form \cite{CeNi2}, with the lattice parameter
practically identical (within 3\%) to that of CeMnNi$_{4}.$ This proves that
the lattice parameter of the latter is defined by the Ce-Ni interaction.
 After one Ce is substituted by a Mn with its
30\% smaller metal radius, Mn appears in a cage much larger than is needed for
normal metallic bonding. Indeed, known Mn-Ni binaries (MnNi, MnNi$_{3})$ are
characterized by the Ni-Mn bonds of the order of 4.8 $a_{B},$ compared to
nearly 5.5 $a_{B}$ in CeMnNi$_{4}.$ Thus, Mn in CeMnNi$_{4}$ is a
\textquotedblleft rattling\textquotedblright\ ion, similar, for example, to La
rattling in thermoelectric skutterudites. This anomalously large distance from
Mn to its nearest neighbors explains why the Mn bands in CeMnNi$_{4}$ are so narrow.

Even a cursory glance at the density of states (Fig. \ref{OPTdos}) and
especially at the band structure (Fig. \ref{OPTbs}) of the ferromagnetic
CeMnNi$_{4}$ reveals that despite the nearly-integer magnetic moment it could
not be farther from a half metal. What is actually happening is that in
both spin channels the Fermi level, rather accidentally, falls inside a deep
pseudogap (about 0.3 eV wide), thus making this material more a semimetal than
half metal (except that in a classical semimetal, like Bi, there is at least a
direct gap, although the valence band and the conductivity bands have a small
indirect overlap, whereas in CeMnNi$_{4}$ there is no gap at all). The DOS at
the Fermi level is $N_{\uparrow}=0.85$ states/eV.formula, $N_{\downarrow
}=1.16$ states/eV.formula, corresponding to an electronic specific heat
coefficient of 4.7 mJ/mol.K$^{2},$ or 0.8 mJ/g-atom.K$^{2}.$ This is a very
small DOS, characteristic rather of doped semiconductors than of metals.

\begin{figure}[tbp]
\centerline{\epsfig{file=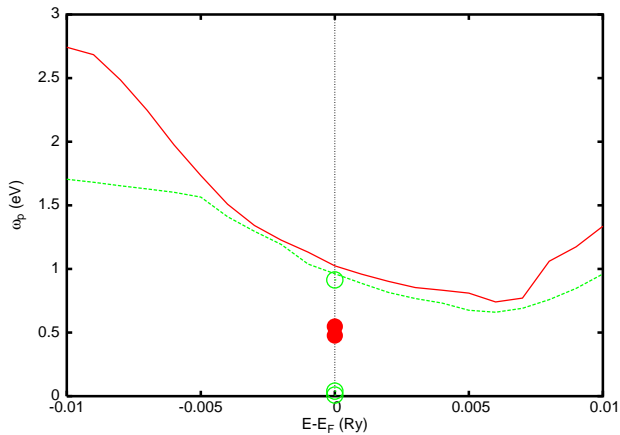,width=0.95\linewidth,angle=0,clip=}}
\caption{The plasma frequencies of ferromagnetic CeMnNi$_{4}$ in the
optimized structure. Green dashed (red solid) lines show the spin-up (spin-
down) components. Symbols show a band decomposition at the Fermi level: 
open: spin-up; filled: spin-down. (color online)}
\label{wp}
\end{figure}
\begin{figure}[tbp]
\centerline{\epsfig{file=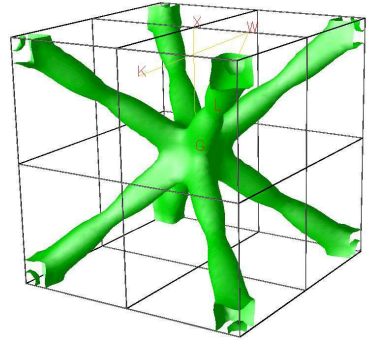,width=0.95\linewidth,angle=0,clip=}}
\centerline{\epsfig{file=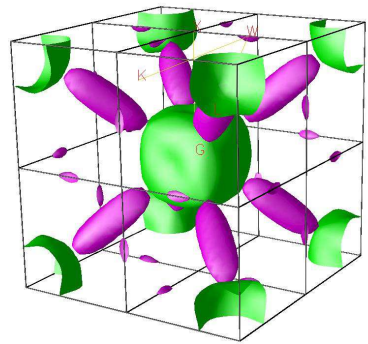,width=0.95\linewidth,angle=0,clip=}}
\caption{The Fermi surfaces of CeMnNi$_4$ for spin-up (top) and spin-down 
(bottom). Only one band is shown for the spin up and two for the spin down. 
Other bands create just barely noticeable Fermi surface pockets.
In order to produce a large number of eigenvalues I used LMTO bands for
this plot; I have verified that the difference beween LMTO and LAPW bands 
is too small to be visible on the scale of this figure.
(color online)}
\label{FS} \end{figure}
Note that the corresponding spin polarization of the DOS is $-16\%,$ far from
the observed 66\%\cite{1}. Of course, one has to keep in mind that the Andreev
reflection is sensitive only to the transport spin polarization, and likely, given
the high resistivity of current samples, to the diffusive transport spin
polarization\cite{PCAR}. Let me remind the reader that the latter can be
expressed in terms of the spin-dependent contribution to the plasma frequency,
$P_{diff}=(\omega_{p\uparrow}^{2}-\omega_{p\downarrow}^{2})/(\omega
_{p\uparrow}^{2}+\omega_{p\downarrow}^{2}).$ Should the Fermi velocities for
the two spin channels be drastically different, that could explain the
observed high transport spin polarization. However, direct calculations yield
the opposite result (Fig.\ref{wp}): $\omega_{p\uparrow}=1.07$ eV,
$\omega_{p\downarrow}=1.10$ eV, corresponding to 3\% spin polarization. This
means that the Fermi velocities are very close for both spins and actually
relatively small for a typical transition metal: $v_{F\uparrow}=2.1\times
10^{7}$ cm/sec, $v_{F\downarrow}=1.9\times
10^{7}$ cm/sec. The message here is that the low
DOS occurs not because of light electrons, but because of the small Fermi
surfaces. Indeed, only three bands, one for the spin-up and two for the
spin-down channel form noticeable Fermi surface pockets, shown in Fig.
\ref{FS}. This emphasizes again the analogy with semimetals.

While the calculations definitely do not agree with the measured spin
polarization, this does not necessarily mean that either are wrong. The
accepted technique for analyzing Andreev reflection data assumes an equal
barrier strength for both spin channels. As has been pointed out
previously\cite{Igor}, this assumption is not always justified and may change
the results substantially. 

One cannot exclude sample problems either; the
temperature dependence reported in Ref. \onlinecite{1} hints at that. Indeed, the
extremely weak temperature dependence of the resistivity above $T_{C}$ implies
that there are no low-energy excitations (phonons or magnons) 
that could scatter electrons (otherwise one would have a linear $T$ dependence, as 
in the Bloch-Gr\"uneisen formula).  On the
other hand, if such excitations were present below $T_{C}$ but disappeared at
the phase transition, a negative temperature coefficient would be
expected just
near the transition temperature, since the electron-scattering will be
not present above $T_C$. On the other hand, if the resistivity were mainly due to static
defects, the temperature dependence of the resistivity could be explained by a
gradual decrease of the carrier concentration with temperature in the ferromagnetic phase, 
from $T=0$ to $T_C$, which would have then to remain constant above
$T_C$. 
However, the effective 
carrier concentration,
$(n/m)_{eff,}$ is nothing but the plasma frequency expressed in
different units, and, as discussed above,  the plasma frequency in  CeMnNi$_{4}$ 
is much larger in the paramagnetic state \cite{note3}, which would yield
a decrease, not increase of $\rho$ with the temperature, with a resistivity minimum near $T_{C}$
(as, for instance, in
Fe$_{x}$Co$_{1-x}$S$_{2},$ see Ref. \onlinecite{CoS2}). 

On the other hand, the behavior below  $T_C$ is reminiscent of the CMR
manganates and some
magnetic semiconductors, where large residual resistivity is also combined with a rapidly
growing resistivity below $T_C$. The low effective carrier density in  
CeMnNi$_4$ supports this analogy.
However, in CMR materials $T_C$ coincides with a metal-insulator
transition, in most cases resulting in a strong
(orders of magnitude) maximum of resistivity near 
$T_C$, instead of rather flat behavior above $T_C$ in CeMnNi$_4$,
or in even more complicated temperature dependences driven by various
structural transformations.
Nevertheless, 
spatial inhomogenuity and percolation effects,
known to be operative in manganates,
may play an important role in CeMnNi$_4$
 too. All this emphasizes again the unusual
character of this material and calls for further experimental studies.

Let me now summarize the results of the calculations. First, despite the
apparent resemblance to a half metal, CeMnNi$_{4}$ is not one. Its magnetic
moment is simply accidentally nearly integer. Second, CeMnNi$_{4}$ exhibits a
very deep pseudogap at the Fermi level, with the 
DOS dropping to a uniquely low value for
an intermetallic compound. Third, despite the small DOS, the Fermi velocity
is also rather low, which makes CeMnNi$_{4}$ electronically similar to
semimetals. Intriguingly, the calculated electronic structure and transport
 properties offer no obvious explanation of the observed temperature 
dependence of the resistivity, which, unless one is willing to write this
 off as a sample problem, represent a very interesting challenge to the 
theory. Finally, the 
crystal structure is essentially set by the Ce-Ni cage,
with Mn rattling in a cavity
 much larger than what is appropriate for this ion. These rather
unusual characteristics should lead to interesting transport and optical
properties. In particular, last but not least, the similarity to semimetals and
presence of rattling phonon modes should make CeMnNi$_{4}$ a very promising
low-temperature thermoelectric, provided it can be synthesized in a
stoichiometric and defect-free form. On the other hand, by intentionally
introducing defects one can create a material with a very high equilibrium
magnetization and very low resistivity, making it a better soft magnetic
material than the ferrites. Obviously, practical applications in this direction
would require optimizing the material to raise its Curie
temperature to room temperature.

The author acknowledges useful discussions with M. Johannes, D. Khomskii
 and B. Nadgorny.

%\end{multicols}{2}

\begin{thebibliography}{9}
\bibitem {note}Since Ce $f-$states are empty and removed from the Fermi level
one can conjecture that spin-orbit effects are not important. Indeed, I have
performed second-variational calculatiuons with spin-orbit and found the
effect to be negligible. By the same token, one does not expect LDA+U$(f)$
calculations to be any different from the regular LDA, and not even more
relevant, given delocalized nature of the $f-$states in this compound. Out of
curiosity, I have perfored LDA+U caculation in the Around-Mean-Field
approximation, appropriate for a metal like this, and using $U=1.25$ and
$J=0.33$ eV for Ce $f$ (as estimated from the atomic loop in an LMTO program),
and again found practically no change from straight LDA.
\bibitem {noteU} Although Mn d-bands are narrow and very likely correlated,
such a large exchange splitting makes applying LDA+U correction to Mn an 
unnecessary complication: it would just increase the splitting between the 
occupied and unocuupied Mn states without any interesting effects at the Fermi level.
\end{thebibliography}
\end{document}